\documentclass[aps,prb,twocolumn,superscriptaddress,showpacs,amsmath,amssymb,notitlepage]{revtex4-1}
\usepackage{graphicx}
\usepackage{amsmath}
\usepackage{amssymb}
\usepackage{float}

\usepackage[dvipsnames]{xcolor}

\usepackage{cancel,xcolor}
\usepackage{indentfirst}
\usepackage{CJKulem}
\usepackage{soul,xcolor}
\setstcolor{red}

\usepackage{ulem}
\usepackage{cancel}
\usepackage[pdfpagemode=UseNone,pdfstartview=FitH,colorlinks=true,linkcolor=blue,urlcolor=blue,anchorcolor=blue,citecolor=blue]{hyperref}

\newcommand{\beq}{\begin{equation}}
\newcommand{\eeq}{\end{equation}}
\newcommand{\beqa}{\begin{eqnarray}}
\newcommand{\eeqa}{\end{eqnarray}}

\def\Aa2#1{\textcolor{magenta}{#1}}

\def\Aa1#1{\textcolor{blue}{#1}}

\def\pra#1{{ Phys.\ Rev. A\/} {\bf#1}}
\def\prb#1{{ Phys.\ Rev. B\/} {\bf#1}}

\def\prl#1{{ Phys.\ Rev.\ Lett.} {\bf#1}}

\def\nat#1{{ Nature} {\bf#1}}

\def\nap#1{{ Nat. Phys.} {\bf#1}}
\def\nac#1{{ Nat. Commun.} {\bf#1}}

\def\napt#1{{ Nat. Photonics} {\bf#1}}

\begin{document}

\title{Topological states and interplay between spin-orbit and Zeeman interactions in a spinful Su-Schrieffer-Heeger nanowire}


\author{Zhi-Hai\!  Liu }
\email{zhihailiu@pku.edu.cn}
\affiliation{Beijing Key Laboratory of Quantum Devices, Key Laboratory for the Physics and Chemistry of Nanodevices, and Department of Electronics, Peking University, Beijing 100871, China}

\author{O. Entin-Wohlman}
\affiliation{School of Physics and Astronomy, Tel Aviv University, Tel Aviv 69978, Israel}

\author{A. Aharony}
\affiliation{School of Physics and Astronomy, Tel Aviv University, Tel Aviv 69978, Israel}

\author{J. Q. You}
\affiliation{Interdisciplinary Center of Quantum Information and Department of Physics, Zhejiang University, Hangzhou 310027, China}

\author{H. Q. Xu}
\email{hqxu@pku.edu.cn}
\affiliation{Beijing Key Laboratory of Quantum Devices, Key Laboratory for the Physics and Chemistry of Nanodevices, and Department of Electronics, Peking University, Beijing 100871, China}
\affiliation{Beijing Academy of Quantum Information Sciences, Beijing 100193, China }

\begin{abstract}
The interplay between the spin-orbit and Zeeman interactions acting on a spinful Su-Schrieffer-Heeger model is studied based on an InAs nanowire subjected to a periodic gate potential along the axial direction.  It is shown that a nontrivial topological phase can be achieved by regulating the confining-potential configuration. In the absence of the Zeeman field,  we prove that  the topology of the chain is not affected by the Rashba spin-orbit interaction due to the persisting chiral symmetry. The energies of the edge modes can be manipulated by varying the magnitude and direction of the external magnetic field. Remarkably, the joint effect of the two spin-related interactions leads  to novel edge states that appear in the gap formed by the anti-crossing of the bands of a finite spinful dimerized chain, and can be merged into the bulk states by tilting the magnetic-field direction.

\end{abstract}


\date{\today}
\maketitle

\section{Introduction}

 A foremost feature of topological insulators is the appearance of metallic edge (or surface) modes inside the insulating bulk energy band-gap~\cite{Moore2010,Hasan2010}.
  These zero-energy edge states, which reside at the boundary of a topological system \cite{Ryu}, are protected by the chiral symmetry of the bulk system.
  In reality this symmetry is often broken. It is therefore of great importance to investigate the bulk-edge relationship under such circumstances.

The simplest  model which captures the topological phase transition as a function of the Hamiltonian parameters  is the one-dimensional Su-Schrieffer-Heeger (SSH) model~\cite{Su1979}. This prototype offers an excellent platform for simulating more intricate topological systems~\cite{Rice1982,Li2014,Xie2019}, and exploring physical effects caused by non-Hermitian Hamiltonians~\cite{Zhu2014,Lieu2018,Yao2018,Kunst2018,Chen2019}. Experimentally, besides explaining the physics of polyacetylene~\cite{Rice1982},  topological phases as found in the SSH model have been realized in certain synthetic dimerized systems, such as optical superlattices~\cite{Lang2012},  ultracold atoms~\cite{Atala2013,Leder2016,Meier2016},  micro-pillar and other photonic crystals~\cite{Parto2018,Jean2017,Solnyshkov2016,Whittaker2019},  atomically engineered superlattices~\cite{Groning2018},  and graphene nanoribbons \cite{Rizzo2018,Drost2017}.  However, studies of the effect of time-reversal  symmetry breaking on a fermionic dimerized chain are still rather scarce. Even more surprising is the confusion in the literature regrading the effects of the spin-orbit interaction.
Spin-orbit interactions are of paramount importance in developing topological insulators and superconductors~\cite{Kane2005,Sau2010}. Several attempts to investigate the impacts of spin-orbit interactions  on the topological features of the SSH model have been reported, see for instance
 Refs.~\onlinecite{Yan2014,Bahari2016}. Regrettably, these papers fail to incorporate correctly the time-reversal symmetry  of spin-orbit interaction in their model Hamiltonians and, in fact, consider the effect of a Zeeman field,  which breaks time-reversal symmetry.
The correct inclusion of the spin-orbit-induced spin flips in tunneling Hamiltonians is via the
(time-reversal symmetric) Aharonov-Casher effect \cite{AC1984}, which appears as
phase factors that dominate the tunneling amplitudes of the dimerized system~\cite{Yao2017,Shahbazyan1994}.


\begin{figure}
\centering
\includegraphics[width=0.46\textwidth]{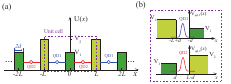}
\caption{(color online) (a)  A  semiconductor wire subjected to  gate potentials $V_{1}$ and $V_{2}$ arranged periodically along the axial direction $x$, to form a double quantum-dot chain. The length of the unit cell is $2L$, and $2d$ is the width of each potential wall. (b) The local potentials $V_{\rm qd,1}(x)$ and $V_{\rm qd,2}(x)$ that define two quantum dots (QD1 and QD2)  within each unit cell, and the lowest orbital on each dot.  Our analysis also applies  for the case  $V_{1}>V_{2}$ (not shown).}
\label{Fig1}
\end{figure}


Recently, the topological phase of the spinful Su-Schrieffer-Heeger configuration
has been approached experimentally~\cite{Whittaker2019,Zhang2017}, by studying the edge modes of a photonic dimerized chain lacking time-reversal symmetry.
In particular, Ref.~\onlinecite{Whittaker2019} which studies topological modes in  micro-pillar lattices (the bosonic analogue of the SSH model) relates the breaking of time-reversal symmetry to a $k$-dependent effective magnetic field acting on the polarization of the  photons.  It is then observed that the edge states do not reside in the mid-gap, as predicted for bosonic systems without chiral symmetry~\cite{Grusdt2013}.
 The analysis of edge modes in  a fermionic topological system where time-reversal and chiral symmetries are not always obeyed thus seems to be quite timely.

\section{Model and  effective Hamiltonian}
  We  present a study of the joint effect of
spin-orbit interaction (SOI) and an external magnetic field on the topological features of  an electron dimerized chain. Our analysis is based on a realistic
Hamiltonian of an InAs nanowire, rendering the conclusions amenable to experimental verifications.  Specifically, the sublattice degrees of freedom of the chain are represented by the localized orbitals of two quantum dots defined by the periodically-arranged gate potentials along the wire axis, as shown in Figs.~\ref{Fig1}(a) and \ref{Fig1}(b).
 Under the interplay between the Zeeman  and  spin-orbit interactions, the strategy in constructing a trackable model for the semiconductor nanowire is based on  the Bloch spectrum of its full  Hamiltonian which is used  to obtain a discretized form, amenable to the analysis of topological features.  This approach, whose validity is demonstrated below, allows us to  determine  the specific form  of the spinful discrete Hamiltonian, for the gated InAs nanowire.

 Assuming that the magnetic field is applied in the $y-z$ plane  along $\hat{\boldsymbol{\varsigma}}$, $\mathbf{B}=B\{0,\cos(\theta),\sin(\theta)\}\equiv B\hat{\boldsymbol{\varsigma}}$,   and the spin-orbit interaction is of the Rashba type, yielding an effective magnetic field along $y$, the Hamiltonian of the InAs nanowire reads~\cite{Liu2018b}
\begin{align}
 H=\frac{p^{2}}{2m_{e}}-V^{}_{\rm c}+U(x)+\alpha p \sigma^{}_{y}+\frac{\Delta_{\rm z}}{2}\sigma^{}_{\varsigma}\ .
 \label{H-0}
 \end{align}
 Here, $m_{e}$ is the electron's effective mass,  $p=-i\hbar\partial/\partial x$ is the momentum, $V^{}_{\rm c}$ denotes the off-set chemical potential, $U(x)$ is the periodic gate potential, $\alpha$ corresponds to the strength of the Rashba SOI, and $\Delta_{\rm z}=g\mu_{\rm B}B$ is the Zeeman splitting, with  $g$ being the Land\'{e} factor ($\mu_{\rm B}$ is the Bohr magneton). The spin-orbit interaction can also be quantified by the length $x_{\rm so}=\hbar/(m^{}_{e}\alpha)$.  In Eq. (\ref{H-0}), $\sigma^{}_{\varsigma}=\hat{\boldsymbol{\sigma}}\cdot\hat{\boldsymbol{\varsigma}}$, with the Pauli matrices $\hat{\boldsymbol{\sigma}}=(\sigma_{x},\sigma_{y},\sigma_{z})$.  The orbital effect of the magnetic field
 is ignored due to the strong confining potential along the direction normal to $\mathbf{B}$, which enables the averaging out the vector-potential terms in the Landau gauge \cite{Liu2018b}.
The periodic  potential creates, in each unit cell, two quantum dots (QD1 and QD2), confined by the potentials
$V_{{\rm qd},1}(x)$ and $V_{{\rm qd},2}(x)$, such that
\begin{align}
V^{}_{{\rm qd},1}(x)&=V_{1}\Theta(d+ x)+V_{2}\Theta(d-L- x)\
\label{vqd12}
\end{align}
with  $-L\leq x\leq L$, where $V^{}_{{\rm qd},2}(x) =V^{}_{{\rm qd},1}(-x)$  and $2L$ is the length of the unit cell [$\Theta(x)$ is the Heaviside function].  $V_{1,2}$ are two separate potential barriers (see  Figs. \ref{Fig1}),  whose width is $2d$.
 All  the calculations below are based on a realistic  InAs nanowire,  for which $m^{}_{e}=0.023m^{}_{0}$ ($m_{0}$ is the free electron mass), the lattice constant is
 $2L=120$~nm, the Land\'{e} $g$ factor is $g=15$~\cite{Winkler2003}, the spin-orbit length is about $x^{}_{\rm so}=180$~nm~\cite{Scherubl2016},  and  the  gate potentials,  of widths $2d=20$ nm, obey $V^{}_{1}+V^{}_{2}=40$~meV~\cite{Supplement0}.



\begin{figure}
\centering
\includegraphics[width=0.48\textwidth]{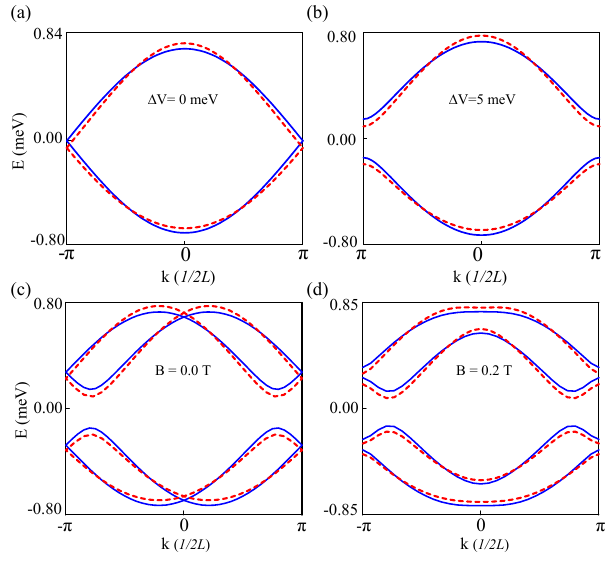}
\caption{(color online)
(a)-(b) The energy spectra of the spinless InAs double quantum-dot chain versus the wave vector $k$ for different potential differences $\Delta V$.  (c)  The Bloch spectrum of  the spinful InAs chain at zero Zeeman field ($B=0$), for  $\Delta V=5.0$~meV and $x^{}_{\rm so}=180$~nm. (d) The same as (c), for B=0.2~T tilted by $\theta=0.5\pi$ in the $y-z$ plane. The  dashed (red) curves depict  the  energy levels of the  Bloch spectrum, calculated from the  continuous Hamiltonian (\ref{H-0}), and  the solid (blue) curves are derived from the discrete  Hamiltonian (\ref{T-B}).  }
\label{Fig2}
\end{figure}

The lowest Zeeman-split levels of the QD's  are used to construct
a discrete Hamiltonian of the double quantum-dot chain
\begin{align}
H^{}_{\rm T}=&\sum_{n}\left(a^{\dagger}_{n}t^{}_{\rm in} b^{}_{n}+a^{\dagger}_{n+1}t^{}_{\rm ex}b^{}_{n}+\rm{H.c.}\right)\nonumber\\
&\hspace{-0.4cm}+\frac{\Delta_{\rm z}}{2}\sum_{n}\left(a^{\dagger}_{n} \sigma^{}_{ z} a^{}_{n}+b^{\dagger}_{n} \sigma^{}_{ z} b^{}_{n}\right)\ ,
\label{T-B}
\end{align}
where $n$ is the index of the unit cell, $a^{}_{n}=\{a^{}_{n\Uparrow}, a^{}_{n\Downarrow}\} $  and $b^{}_{n}=\{b^{}_{n\Uparrow}, b^{}_{n\Downarrow}\} $ are the spinors  for the quasi-spin states on  QD1 and  QD2, respectively. Under the joint effect of the spin-orbit and Zeeman interactions, the intra/inter-cell tunneling amplitudes $t^{}_{\rm in/ex}$ are matrices in spin space,
\begin{align}
t^{}_{\rm in/ex }= t^{}_{{\rm in/ex}, 0} \exp\left[i\varphi^{}_{{\rm in/ex} } (\hat{\mathbf{v}}^{}_{\rm in/ex}\cdot\hat{\boldsymbol{\sigma}})\right]\ .
\label{TAS}
\end{align}
The phase factors $\varphi^{}_{\rm in/ex}\propto L/x_{\rm so}$  and the unit vectors $\hat{\mathbf{v}}^{}_{{\rm in/ex}}=\{\sin(\vartheta^{}_{\rm in/ex}),0,\cos(\vartheta^{}_{\rm in/ex})\}$ are  determined by the relative angle $\theta$ between the directions of the  Zeeman  field and the one induced by the SOI~\cite{Liu2018}.
 Details are given in Appndix \ref{S-1}. Notably, the spin space for defining the Pauli matrices $\sigma^{}_{x,y,z}$ in Eqs.~(\ref{T-B}) and (\ref{TAS})  consists of the two  lowest Zeeman splitting levels on each dot, and does not coincide with that defining the spin operators in Eq.~(\ref{H-0}).  Therefore, the Zeeman interaction  in Eq.~(\ref{H-0})  appears as a $\sigma^{}_{z}$ term in Eq.~(\ref{T-B}).

 In the absence of the  Zeeman and spin-orbit interactions, i.e., $\Delta^{}_{\rm z}=0$ and $\varphi^{}_{\rm in/ex}=0$, the spinless version of $H^{}_{\rm T}$ describes the well-known Su-Schrieffer-Heeger model \cite{Su1979,Delplace2011}
$
H^{}_{\rm T,0}=\sum_{n}(t^{}_{\rm in,0} a^{\dagger}_{n}b^{}_{n}+t^{}_{\rm ex,0}a^{\dagger}_{n+1}b^{}_{n}+\rm{H.c.}),
$
with $a^{}_{n}$  and $b^{}_{n}$  representing the annihilation  operators for the two-site unit cell.   As is well known, this model  leads to a topological transition depending  on the degree of dimerization, denoted as $t^{}_{-}$ with
\begin{align}
t^{}_{\pm}= t^{}_{\rm ex,0 }\pm t^{}_{\rm in,0 }\ ,
\label{DT}
 \end{align}
 which for the InAs nanowire $t^{}_{-}$ is determined by  $\Delta V=V_{1}-V_{2}$.  Figures \ref{Fig2}(a) and \ref{Fig2}(b)
exhibit comparisons between the spinless Bloch spectra obtained from the ``exact" Hamiltonian in Eq.~(\ref{H-0}), and from the approximated one in Eq.~(\ref{T-B}), which show that the spectra of the latter follows those of  the former quite faithfully~\cite{Supplement2}.
A further support  for the compatibility of our tight-binding description for a realistic InAs nanowire
emanates from the fact that
 the band-gap closes at the critical point  $\Delta V=0$. Specifically, the ``exact'' Bloch spectrum of the synthetic lattice is derived   exploiting periodic boundary conditions of the Bloch functions in a unit cell, see Appendix~\ref{S-2}.  Moreover, using the cell-periodic Bloch functions
 $u^{}_{k}(x)=u^{}_{k}(x+2L)$, one obtains the
 Zak phase  $
\varphi^{}_{\rm zak }=i\int^{\pi}_{-\pi} \int^{L}_{-L} u^{\dagger}_{k }(x)\frac{\partial}{\partial k}u^{}_{k } (x)dx dk$~\cite{Zak1989,Hatsugai2006}.
The integration is carried out by
the discretization method introduced in Ref.~\onlinecite{Fukui2005}.
It is found that the Zak phase of each Bloch band depends on the confinement configuration such that
$
 \varphi^{}_{\rm zak}=0$ for  $\Delta V<0$, and  $\varphi^{}_{\rm zak}=\pi$ for $ \Delta V >0$,
 in full agreement with the one derived from the tight-binding description of the SSH model~\cite{Delplace2011}, based on the dependence of the dimerization degree  on the potential difference.


 The Bloch bands of the spinful Hamiltonian (\ref{H-0}) are displayed in Figs. \ref{Fig2}(c) and \ref{Fig2}(d) by the dashed curves. The ones belonging to the discretized version,  Eq.~(\ref{T-B}), are shown as solid curves. Again, a good agreement between the two spectra is obtained. Note in particular the closing of the gap at $k=0$ and $k= \pm \pi $ for zero magnetic field, similarly to the spinless SSH model. This results from the chiral symmetry of the Rashba interaction (see below).

 \section{Topological edge modes for broken  chiral symmetry}

\begin{figure}
\centering
\includegraphics[width=0.42\textwidth]{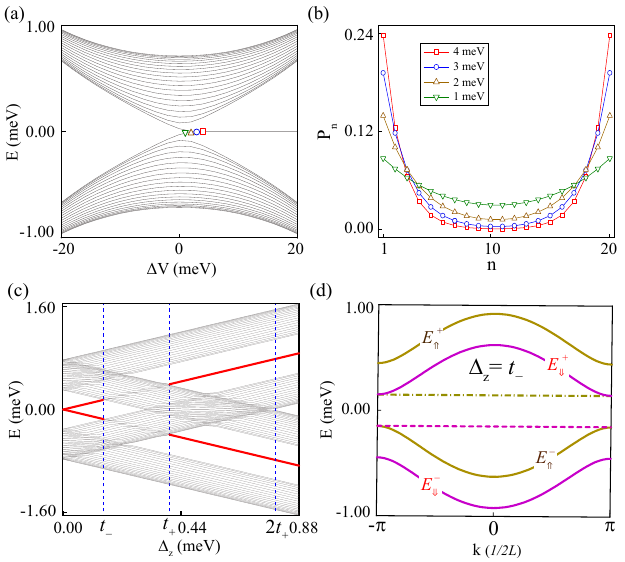}
\caption {(color online) (a) The energy spectrum of a finite spinless chain versus  $\Delta V$. The spatial density distributions of one spinless edge mode, $P_{n}$,  for different values of $\Delta V$, as indicated by the symbols in panel (a),  are displayed in (b). (c) The energy spectrum of a  finite chain of 20 sites versus $\Delta^{}_{\rm z}$  for $\Delta V=10$~meV, $\theta=0.5\pi$, and zero spin-orbit coupling. The  emergence of the Zeeman-split edge states is  indicated by the thick (red) lines. (d) The Bloch  spectrum of the semiconductor chain as a function of $k$ with $\Delta^{}_{\rm z}=t^{}_{-}$,  $\Delta V=10$~meV, $\theta=0.5\pi$, and zero spin-orbit coupling. Besides, the energy values of the spin-up  and spin-down edge states are indicated by the (yellow) dash-dotted and (purple) dashed lines, respectively.
  }
\label{Fig3}
\end{figure}

 By analogy with the spinless SSH model,
the nontrivial topological phase  is  characterized by the appearance, at $\Delta V>0$,  of two zero-energy edge modes in a finite chain of $N$ sites with open boundary conditions, see Figs.~\ref{Fig3}(a) and \ref{Fig3}(b).
 In the thermodynamic limit ($N\rightarrow \infty$), these two zero-energy  edge modes are separately located on either side of the chain  and  hence  each belongs to a different sublattice~\cite{Ryu,Asboth2016}.
 For an open short chain comprising 20 sites, the separated edge modes are coupled to each other; the hybridized  edge modes can be written as  $|E_{\pm }\rangle=\sum^{20}_{n=1}(A^{ }_{n}a^{\dagger }_{n}\pm B^{ }_{n}b^{\dagger}_{n})|0\rangle$,  where  $|0\rangle$ is the vacuum state  and $A^{}_{n}$ and $B^{}_{n}$  are  the probability amplitudes on the two  sites of the $n$-th unit cell. In this case, the spatial density distribution  of the two edge modes at the $n$th site is $P_{n}=|A^{}_{n}|^{2}+|B^{}_{n}|^{2}$ and it depends on the potential  difference $\Delta V$. As seen in Fig.~\ref{Fig3}(b), the larger $\Delta V$ is, the more localized is  the edge state, due to the accompanying  enhanced dimerization strength~\cite{Nevado2017}.

%



Interestingly, the presence of the spin-orbit coupling does not cause qualitative changes in this picture. At zero Zeeman field,  we find zero-energy edge modes. [Note that the energy spectrum  of an open spin-orbit-active chain is a doubled copy of that of the spinless  chain  shown in Fig.~\ref{Fig3}(a).]  In fact, these zero-energy modes are protected by the chiral symmetry of the discrete Hamiltonian (\ref{T-B}) for   $\Delta^{}_{\rm z}=0$: $\mathcal{C}^{}_{}H^{}_{\rm  T}\mathcal{C}_{}^{-1}=H^{}_{\rm  T}$,  where  the chiral symmetry operator is $\mathcal{C}=\mathcal{T}\mathcal{P}$.
 Here,
 $\mathcal{T}^{}_{}$ is the time-reversal operator,  and
$\mathcal{P}$ is the particle-hole operator, which for a spinful bipartite lattice is defined by
 $\mathcal{P}^{}_{}a^{}_{\Uparrow/\Downarrow}\mathcal{P}_{}^{-1}=\pm a^{\dagger}_{\Downarrow/\Uparrow}$ and $\mathcal{P}^{}_{}b^{}_{\Uparrow/\Downarrow}\mathcal{P}_{}^{-1}=\mp b^{\dagger}_{\Downarrow/\Uparrow}$.

The Zeeman interaction which breaks time-reversal symmetry lifts the chiral symmetry.
  In the absence of  the spin-orbit interaction (i.e., for $\varphi^{}_{\rm in/ex}=0$),
the spinful Hamiltonian $H^{}_{\rm T}$
is separated into two independent models for each spin projection, $H^{}_{\rm T}=H^{}_{\Uparrow}+H^{}_{\Downarrow}$ with
\begin{align}
H^{}_{\tau=\Uparrow/\Downarrow}=&\sum_{n}\left(t^{}_{\rm in,0}a^{\dagger}_{n\tau}b^{}_{n\tau}+t^{}_{\rm ex,0}a^{\dagger}_{n+1\tau}b^{}_{n\tau}+\rm{H.c.}\right)\nonumber\\
&\hspace{-0.4cm}\pm\frac{\Delta_{\rm z}}{2}\sum_{n}\left(a^{\dagger}_{n\tau}  a^{}_{n\tau}+b^{\dagger}_{n\tau}  b^{}_{n\tau}\right)\ .
\label{T-B-S}
\end{align}
Structurally, $H^{}_{\Uparrow /\Downarrow}$ can be mapped onto an effective SSH model with a finite on-site energy $\pm \Delta^{}_{\rm z}/2$.  It follows that in principle  there exist two  edge  states in each spin sector for $t^{}_{-}> 0$, see Eq.~(\ref{DT}).   However,
 there is a caveat: since each spin-polarized edge mode  can merge into the bulk states pertaining to the other spin direction, which happens for  $t^{}_{-}\leq\Delta^{}_{\rm z}\leq t^{}_{+}$,
the available Zeeman
splittings for which
spin-polarized edge modes can appear is restricted, as shown in Fig.~\ref{Fig3}(c).

To further illustrate this point, we examine the Bloch spectrum in the absence of the SOI,
\begin{align}
E^{\pm }_{\Uparrow} (k)=& \frac{\Delta^{}_{\rm z}}{2}\pm \sqrt{t^{2}_{\rm in,0}+ t^{2}_{\rm ex,0} +2t^{ }_{\rm in,0}t^{ }_{\rm ex,0}\cos k} \label{BES} \\
E^{\pm }_{\Downarrow } (k)=& -\frac{\Delta^{}_{\rm z}}{2} \pm \sqrt{t^{2}_{\rm in,0}+ t^{2}_{\rm ex,0} +2t^{ }_{\rm in,0}t^{ }_{\rm ex,0}\cos k}\ .\nonumber
\end{align}
For a zero magnetic field, the energies of the edge states vanish; when this field is active and the edge states become spin polarized, their respective energies are
$\varepsilon^{}_{\Uparrow/\Downarrow}=\pm \Delta^{}_{\rm z}/2 $.
The  energy of the  spin-up  edge state, $\varepsilon^{}_{\Uparrow}$, intersects the upper Bloch band $E^{+}_{\Downarrow}(k)$ for $t^{}_{-}\leq \Delta^{}_{\rm z}\leq t^{}_{+}$. This conclusion  also pertains to the intersection between $\varepsilon^{}_{\Downarrow}$ and $E^{-}_{\Uparrow}(k)$, as depicted in Fig.~\ref{Fig3}(d). As a result, the intersection is reflected by the coalescence of the bulk and   edge states, as seen in Fig.~\ref{Fig3}(c).


\begin{figure}
\centering
\includegraphics[width=0.46\textwidth]{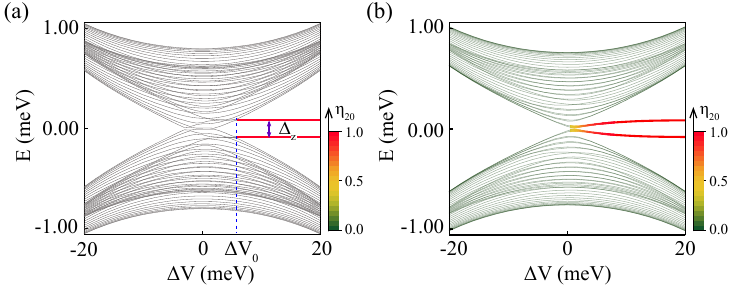}
\caption {(color online)  (a) The energy spectrum of a finite spinful InAs chain versus $\Delta V$ for  $B=0.2$~T, $\theta=\pi/2$ and in the absence of the SOI. The edge states appearing in the band-gap are indicated by the thick (red) lines.  (b) The same as (a),  but in the presence of the SOI. The  coefficient $\eta^{}_{20}$   characterizes the degree of confinement of the edge states  in a $20$ site chain. }
\label{Fig4}
\end{figure}

 For a Zeeman energy such that  $\Delta^{}_{\rm z}<t^{}_{+}$,
  the  edge states appear above a certain value of the dimerization strength,   $ t^{}_{-}>\Delta^{}_{\rm z}$, as reflected by    the threshold value, denoted   $\Delta V^{}_{0}$ in Fig.~\ref{Fig4}(a).  This condition disappears in the presence of the SOI,  as seen in Fig.~\ref{Fig4}(b), indicating that there is no further requirement on the dimerization strength in the nontrivial topological phase, i.e., $\Delta V>0$.   This feature can  be understood by referring to the  spin-orbit-active qubit   discussed in Ref.~\onlinecite{Li2013}.
 The Land\'{e} $g$-factor in the presence of the SOI, and consequently $\Delta^{}_{\rm z}$, are reduced  by
  a SOI-dependent factor  $f=\exp[-x_{0}^{2}/x^{2}_{\rm so}]$~\cite{Li2013}, where  $x^{}_{0}$   is the localization length of the edge mode. In fact, $x^{}_{0}$  is correlated with the  probability density distribution of the edge state $\widetilde{P}^{}_{n}=\sum^{}_{\tau=\Uparrow,\Downarrow}[|A^{}_{n,\tau}|^{2}_{}+|B^{}_{n,\tau}|^{2}_{}]$,
  where $A^{}_{n,\tau}$ and $B^{}_{n, \tau}$ are the amplitudes of the quasi-spin states on the QD1 and QD2,  respectively. One may characterize the degree of  confinement to the edges  by introducing the measure  (pertaining to a chain comprising $20$ sites)
  $\eta^{}_{20}=\sum^{3}_{n=1} \widetilde{P}_{n}+\sum^{20}_{n'=18}\widetilde{P}^{}_{n'}$.
As portrayed in Fig.~\ref{Fig4}(b), $\eta_{20}^{}$ is reduced as  $\Delta V$ is decreased, implying that the localization length is lengthened.
As  large $x^{}_{0}$'s are realized at  small $\Delta V$'s   this explains the results in that regime.

\begin{figure}
\centering
\includegraphics[width=0.48\textwidth]{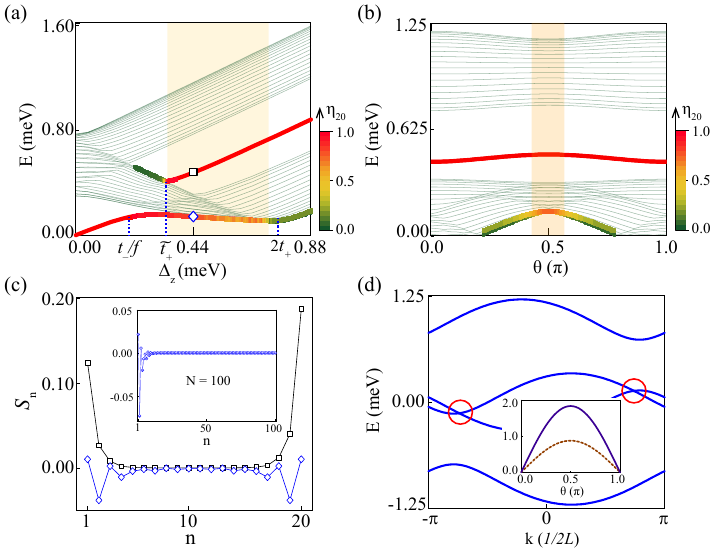}
\caption{(color online)  (a) The positive energy spectrum of a finite spin-orbit-active chain of 20 sites versus the Zeeman splitting,  with  $\Delta V= 10$~meV and $\theta=0.5\pi$.  (b) The positive energy spectrum of the same chain  versus the tilting angle $\theta$ of ${\bf B}$ in the $y-z$ plane  for $\Delta^{}_{\rm z}=0.44$~meV and $\Delta V=10$~meV. Also shown are the (yellow) shadowed  regions where there exist four pairs of edge modes whose degree of confinement is characterized by the coefficient $\eta^{}_{20}$. (c) The  spin-density distributions, $S^{}_{n}$, of the two edge modes marked by a square and a diamond in panel (a). Moreover,  for a   discrete chain comprising of 100 sites, the spin-density distribution of the new edge mode [corresponding to the diamond in panel (a)] is further exemplified in the inset.
(d) The Bloch spectrum of the spin-orbit-active chain versus $k$ for $\Delta V=10$~meV, $\Delta^{}_{\rm z}=0.44$~meV, and $\theta=0$. In this case, the anti-crossing gap accommodating the spin-mixed edge states completely disappears, as indicated by the red circles. In addition,  the inset shows the dependence  of the two spin-flip tunneling amplitudes   (in meV units)  on $\theta$, with the (purple) solid and (yellow) dashed curves representing the inter- and intra-cell tunneling amplitudes, $|t^{\prime}_{\rm  ex}|$  and $|t^{\prime}_{\rm in}|$, respectively. }
\label{Fig5}
 \end{figure}

In the presence of a Zeeman field, the energies of the
two doubly-degenerate edge modes shown in Figs.~\ref{Fig4}(a) and \ref{Fig4}(b)
  are split and localized symmetrically around zero energy, since the Hamiltonian obeys particle-hole symmetry, $P_{}H_{\rm T}P^{}_{-1} = H^{}_{\rm T}$.  Therefore, in order to facilitate the following analysis, we only focus on the variations of the upper energy levels  under different conditions.
For a fixed  value of  $\Delta V$ the  precondition for the  appearance of the  spin edge states is modified  under the joint effect of  the  spin-orbit and Zeeman interactions.
Because the energies of the Zeeman-split edge modes are modified by the SOI-induced factor $f$, the available range of $\Delta^{}_{\rm z}$ near zero magnetic field   which supports edge modes is changed to $\Delta_{\rm z} <t_{-}/f$,  with   $f\simeq0.8$ for $\Delta V=10$~meV, see Fig.~\ref{Fig5}(a).

The condition for the emergence of  edge states at large magnetic fields, $\Delta^{}_{\rm z}>t_{+}$,  is also modified in the presence of the SOI.  While in the absence of SOI, the edge modes are the  eigenvectors of $\sigma_{z}$ separated by $\Delta^{}_{\rm z}$, they become coupled  by the spin-flip tunneling  matrix elements, $t^{\prime}_{\rm in/ex}$, induced by the SOI [see Eq.~(\ref{TAS})].  The resulting edge states can be explained by confining the discussion to the thermodynamic limit.
 Exploiting
 the orthonormal basis    of $\sigma_{z}$,   the Hamiltonian (\ref{T-B}) is approximated by
\begin{align}H^{}_{\rm T}=\begin{pmatrix}
\frac{\Delta^{}_{\rm z}}{2}&t^{\prime}_{}\\
 t^{\prime}_{}&-\frac{\Delta^{}_{\rm z}}{2}
 \end{pmatrix}
\label{MAT}
\end{align}
  where  $t^{\prime}_{}=\sqrt{|t^{\prime}_{\rm in}|^{2}+|t^{\prime}_{\rm ex}|^{2}}$  represents the spin-flip coupling.
 The two eigenvalues of  Eq.~(\ref{MAT})  are  the modified   energies of the quasi-spin edge modes.   The lower bound for their existence is
  $\widetilde{t}^{}_{+}=\sqrt{t^{2}_{+}-4t^{\prime2}_{}}$. For  $\Delta^{}_{\rm z}< \widetilde{t}^{}_{+}$, the quasi-spin edge modes are merged into the bulk states, as reflected by the sharp decrease  of the confinement coefficient $\eta_{20}^{}$ shown in Fig.~\ref{Fig5}(a).
 The energies of these Zeeman-split edge  states depend on the tilting angle $\theta$ of the magnetic field as shown in Fig.~\ref{Fig5}(b), and hence can be varied  in experiment.

  Remarkably, there appear additional edge modes for Zeeman splittings obeying  $t^{}_{-}/f\leq\Delta^{}_{\rm z}<2t_{+}$, see Fig.~\ref{Fig5}(a), a range which is ``forbidden" in  Fig.~\ref{Fig3}(c).
  These  modes result from the interplay between the Zeeman and spin-orbit interactions.  They reside below  the anti-crossing gap  induced by the  spin-flip tunneling matrix elements in Eq.~(\ref{TAS}): The crossing of the Bloch bands   $E^{-}_{\Uparrow}(k)$ and $E^{+}_{\Downarrow}(k)$ in  momentum space  is lifted  by these spin-flip amplitudes, and a gap (``anti-crossing gap") is open. Indeed, introducing those as perturbations on the energy bands in Eqs.~(\ref{BES}), one observes that the crossing, and simultaneously the gap opened by the spin-flip terms, disappear for $\Delta_{\rm z}>2t_{+}$.
 Hence, we deduce that  these extra  edge modes amalgamate with the bulk states   for    a  Zeeman splitting smaller than the critical value, $ 2t^{}_{+}$. This is reflected by the variation of $\eta^{}_{20}$ shown in Fig.~\ref{Fig5}(a).
  Numerically, a mode (of a 20 site chain) can be considered as an edge state for, say, $\eta^{}_{20}>0.5$, and the mergence  of the new edge states into the bulk is verified  by this criterion, correspondingly.

As compared to the Zeeman-split (higher energy) edge modes depicted in Fig.~\ref{Fig5}(a), the energy eigenvalues of these new edge states are insensitive to the increase of the Zeeman splitting due to the high degree of spin mixing.  For this end,  Fig.~\ref{Fig5}(c) displays  the distributions of the spin density $S^{}_{n}=\frac{1}{2}[|A^{}_{n,\Uparrow}|^{2}+|B^{}_{n,\Uparrow}|^{2}-|A^{}_{n,\Downarrow}|^{2}-|B^{}_{n,\Downarrow}|^{2}]$,  for the two edge modes illustrated in Fig.~\ref{Fig5}(a). Evidently,  the new edge state (marked by the diamond) is distinguished from  the spin-resolved edge mode (indicated by the square)  by  its small  spin-density probability $S_{n}$. Similar to the spatial distribution of the spinless edge modes  shown in Fig.~\ref{Fig3}(b),  the spin-density distribution  of each spinful edge mode  is split into two segments for an open short discrete chain  comprising of 20 sites,  see Fig.~\ref{Fig5}(c).
 Analogously, because  the  coupling   between the  two presumably separated edge modes (located in different sides of the chain) is negligible in the thermodynamic limit,   the extra edge mode can only reside in either side of the chain, as indicated by the  spin-density   distribution $S^{}_{n}$ shown in the inset of Fig.~\ref{Fig5}(c) for $N=100$.

Hence, the total number  of the edge modes can be doubled when $\widetilde{t}^{}_{+}<\Delta^{}_{\rm z}<2t^{}_{+}$, as indicated  in Fig.~\ref{Fig5}(a). However, in contrast to the Zeeman-split edge states,  the spin-mixed edge modes can be merged into the bulk states by tilting  the magnetic field in the $y-z$ plane, as shown in Fig.~\ref{Fig5}(b), due to the dependence of the tunneling amplitudes on the tilting angle, see Eq.~(\ref{TAS}).
 In particular, for  $\theta=0$, the spin of the dimerized chain is conserved, $\vartheta^{}_{\rm in/ex}=0$ and $t^{\prime}_{\rm in/ex}=0$, and the gap opened due to the anti-crossing of the spin-resolved energy levels completely disappears, as shown in Fig.~\ref{Fig5}(d).
\\

\section{Conclusions}

  Specifying to an InAs nanowire with strong Rashba spin-orbit interaction, we have shown that   this interaction by itself  does not affect the topology of a spinful  Su-Schrieffer-Heeger  model,  as opposed  to the significant effects of spin-orbit interaction found  in Refs.~\onlinecite{Bahari2016,Whittaker2019,Yan2014}.   In the presence of an external magnetic field, we find that the  joint effect of the spin-orbit and Zeeman interactions   allows for a plethora of edge modes  which are not protected by chiral symmetry. The edge states can be classified into two groups. The first includes the Zeeman-split edge states, and the second comprises the ones induced by the interplay between the two spin-related interactions appearing at a sufficiently large Zeeman splitting. Interestingly, these new states exist in the gap opened by the anti-crossing of the energy bands and may merge into the bulk by tilting the magnetic-field direction. This remarkable observation can be examined experimentally using  semiconductor nanowires~\cite{Nilsson2009,Mu2021,Jong2019} and graphene nanoribbons~\cite{Groning2018,Rizzo2018}.


\begin{acknowledgments}
This work is supported by the Ministry of Science and Technology of China through the National Key Research and Development Program of China (Grant Nos. 2017YFA0303304 and 2016YFA0300601), the National Natural Science Foundation of China (Grant Nos. 91221202, 91421303, 11874071 and 11934010), the Beijing Academy of Quantum Information Sciences (No. Y18G22), and the Key-Area Research and Development Program of Guangdong Province (Grant No. 2020B0303060001).
 OEW and AA acknowledge support by the Israel Science Foundation (ISF), by the infrastructure program of Israel Ministry of Science and Technology under contract 3-11173, and by the Pazy Foundation.
\end{acknowledgments}

\begin{appendix}

\begin{widetext}
\section{The tight-binding Hamiltonian of the model system}
\label{S-1}

\begin{figure}
\centering
\includegraphics[width=0.7\textwidth]{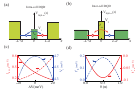}
\caption{ (color online)
  (a)  The spatial distribution of the confining potential $V^{}_{\rm DQD-in}(x)$ that determines the intra-cell tunneling amplitude within the double quantum-dot chain in the main text (see Fig.~\ref{Fig1}). (b) The confining potential $V^{}_{\rm DQD-ex}(x)$ that determines the inter-cell tunneling amplitude. (c) The intra/inter-cell hopping amplitude of an InAs double quantum dot chain, $t^{}_{\rm in/ex,0}$, as a function  of the potential difference $\Delta V=V^{}_{1}-V^{}_{2}$, with the  periodic length $2L=120$~nm and  $d=10$~nm. Here, $t^{}_{\rm in,0}$ ($t^{}_{\rm ex,0}$) is denoted by the solid (dashed) line and the variations of the  chemical potential $V_{\rm c}$ are displayed by the dash-doted curve.  (d)   The intra- and inter-cell spin-flip tunneling amplitudes, $t^{\prime}_{\rm in}$ and  $t^{\prime}_{\rm ex}$, of the double-quantum-dot chain as functions of the tilting angle $\theta$, with the potential difference $\Delta V^{}_{}=-10$~mev,  the  magnetic field strength $B=1.0$~T, and the spin-orbit interaction scaled by $x^{}_{\rm so}\equiv \hbar/(m_{e}\alpha)=180$~nm.}
\label{Figs1}
\end{figure}


%
 Here is outlined the  derivation of the  nearest-neighbor tight-binding (TB) Hamiltonian of our model.
 The derivation is based on the original Hamiltonian, Eq.~(\ref{H-0}) in the main text. The periodic gate potential there, $U(x)$,  is viewed as composed of two alternating quantum dots, QD1 and QD2, as depicted in Fig.~\ref{Figs1}(a) and (b). Each quantum dot obeys the Hamiltonian
 \begin{align}
 H^{}_{\rm 1/2}=\frac{p^{2}}{2m_{e}}+ V^{}_{\rm qd,1/2}(x)+\alpha p\sigma_{y}+\frac{\Delta_{\rm z}}{2}\sigma^{}_{\varsigma},
 \label{H12}
 \end{align}
 where  $\alpha$ is  the strength of the Rashba spin-orbit interaction, $\Delta_{\rm z}$ is the Zeeman splitting, $\sigma^{}_{n}= \hat{\boldsymbol{\sigma}}\cdot \hat{\boldsymbol{\varsigma}}$ with  $\hat{\boldsymbol{\varsigma}}=\{0, \cos\theta, \sin\theta\}$ being the direction vector of the external magnetic field  and  $\hat{\boldsymbol{\sigma}}$ denotes  the Pauli matrices,  $\hat{\boldsymbol{\sigma}}=\{\sigma^{}_{x},\sigma^{}_{y},\sigma^{}_{z}\}$.  The  confining  potentials that define the quantum dots are
 \begin{align}
 V^{}_{\rm qd,1}(x)=\begin{cases}
 V_{1} ~~~~~~~~~~~~~x\geq -d\\
 0~~~~~~~ -L+d\leq x<-d,\\
 V_{2}~~~~~~~~~~~x<-L+d
 \end{cases}~~~~   V^{}_{\rm qd,2}(x)=\begin{cases}
 V_{1} ~~~~~~~~~~~~~x< d\\
 0~~~~~~~ d\leq x<L-d.\\
 V_{2}~~~~~~~~x\geq L-d
 \end{cases}
 \label{lcp}
 \end{align}

 We next solve for the energy eigenstates of the two dots using Eq. (\ref{H12}). Moreover,  in order to preserve the locality of the wave functions, we assume  the electron energy eigenvalue   $E<\min\{V^{}_{1},V^{}_{2}\}$.  It
is expedient to consider the Schr\"{o}dinger equation pertaining to all three regions in Eq. (\ref{lcp}) by introducing
 \begin{align}
 \widetilde{H}^{}_{\zeta=0,1,2}\psi^{}_{\zeta }(x)=E\psi_{\zeta}(x).
 \label{WQD}
 \end{align}
 Here,   $\psi^{}_{\zeta}$ is the corresponding eigenfunction on each portion  and
\begin{align}
\widetilde{H}^{}_{\zeta}=\frac{p^{2}}{2m_{e}}+\alpha p\sigma^{}_{y}+\frac{\Delta^{}_{\rm z}}{2}\sigma^{}_{\varsigma}+V^{}_{\zeta},
\label{LQD}
 \end{align}
  where
 $V_{0}=0$, and $V_{1,2}$ are given in Eqs. (\ref{lcp}).
The eigenfunctions of $\widetilde{H}^{}_{\zeta}$ are spinors,
 \begin{align}
 \psi^{}_{\zeta}(x)=\exp\left(i\gamma^{}_{\zeta}x\right)\begin{pmatrix}
 w_{\zeta}\\
 d_{\zeta}
 \end{pmatrix}.
 \label{BEFs}
 \end{align}
For a fixed value of $E$ there exist four independent solutions to Eq.~(\ref{WQD}), with the corresponding wave vectors  $\gamma^{}_{\zeta, \lambda=1,2,3,4}$  determined by the  quartic equation
  \begin{align}
  \left(\frac{\hbar^{2}{\gamma^{2}_{\zeta}}}{2m_{e}}+V^{}_{\zeta}-E\right)^{2}-\hbar^{2}\gamma^{2}_{\zeta}\alpha^{2}-\Delta_{\rm z}\hbar \gamma^{}_{\zeta}\alpha\cos(\theta)-\frac{\Delta^{2}_{\rm z}}{4}=0.
  \label{wk}
  \end{align}
  The respective spinor components are
 \begin{align}
w^{}_{\zeta,\lambda}= i\hbar\alpha\gamma^{}_{\zeta, \lambda}+i\frac{\Delta^{}_{\rm z}}{2}\cos(\theta),~~~~~~d^{}_{\zeta,\lambda}=\frac{\hbar^{2}\gamma^{2}_{\zeta, \lambda}}{2m_{e}}+\frac{\Delta_{\rm z}}{2}\sin(\theta)+V^{}_{\zeta}-E.
  \label{wkc}
 \end{align}
 In particular, when $\theta=\pi/2$, the expressions for the wave vectors can be derived analytically
 \begin{align}
 \gamma^{}_{\zeta,1/2}=&\pm\sqrt{\frac{2m_{e}}{\hbar^{2}}}\sqrt{E-V_{\chi}+m_{e}\alpha^{2}+\sqrt{m^{2}_{e}\alpha^{4}+2m_{e}\alpha^{2}(E-V_{\chi})+\Delta^{2}_{\rm z}/4}} \nonumber\\
 \gamma^{}_{\zeta,3/4}=&\pm\sqrt{\frac{2m_{e}}{\hbar^{2}}}\sqrt{E-V_{\chi}+m_{e}\alpha^{2}-\sqrt{m^{2}_{e}\alpha^{4}+2m_{e}\alpha^{2}(E-V_{\chi})+\Delta^{2}_{\rm z}/4}}.
 \end{align}
The convergence of the wave function at  $x\rightarrow \infty$ requires the wave vectors to have ${\rm Im}[\gamma^{}_{1}]>0$  for dot 1 and ${\rm Im}[\gamma^{}_{2}]>0$ for dot 2.  The convergence condition for $x\rightarrow -\infty$ is ${\rm Im}[\gamma^{}_{2}]<0$  for dot 1 and ${\rm Im}[\gamma^{}_{1}]<0$ for dot 2.
Then, the energy eigenfunctions of QD1 and QD2 can be written as
\begin{align}
\Psi_{1}(x)=&\sum^{4}_{\lambda=1}c^{}_{0,\lambda}[1-\Theta(x+d)]\Theta(x+L-d)\psi^{}_{0,\lambda}(x)+\Theta(x+d)[c^{}_{1,1}\psi^{}_{1,1}(x)+c^{}_{1,4}\psi^{}_{1,4}(x)]\nonumber\\
&~~~~~~~+[1-\Theta(x+L-d)][c^{}_{2,2}\psi^{}_{2,2}(x)+c^{}_{2,3}\psi^{}_{2,3}(x)]\\
\Psi_{2}(x)=&\sum^{4}_{\lambda=1}v^{}_{0,\lambda}[1-\Theta(x-L+d)]\Theta(x-d)\psi^{}_{0,\lambda}(x)+[1-\Theta(x-d)][v^{}_{1,2}\psi^{}_{1,2}(x)+v^{}_{1,3}\psi^{}_{1,3}(x)]\nonumber\\
&~~~~~~~+\Theta(x-L+d)[v^{}_{2,1}\psi^{}_{2,1}(x)+v^{}_{2,4}\psi^{}_{2,4}(x)],
\end{align}
where $\Theta(x)$ is the Heaviside step function  $\Theta(x)=\begin{cases}
0  ~~x\leq0\\
1 ~~x>0\end{cases}$, the coefficients $c^{}_{\zeta,\lambda}$ and  $v^{}_{\zeta,\lambda}$  are determined by the boundary conditions, i.e., from the continuity  equations at $x^{}_{1/3}=\mp d$ and $x^{}_{2/4}=\mp L\pm d$.
 Using the explicit form for $V^{}_{qd,1}(x)$ in Eq.~(\ref{lcp}), the boundary conditions for the localized wave function on QD1 are \cite{com}
\begin{align}
\lim_{\epsilon \rightarrow 0 }[\Psi_{1}(x_{1}+\epsilon)-\Psi_{1}(x_{1}-\epsilon)]=0 ,~~~~\lim_{\epsilon\rightarrow 0} [\Psi_{1}(x_{2}+\epsilon)-\Psi_{1}(x_{2}-\epsilon)]=0 , \nonumber\\
\lim_{\epsilon \rightarrow 0 }[\Psi'_{1}(x_{1}+\epsilon)-\Psi'_{1}(x_{1}-\epsilon)]=0 ,~~~~\lim_{\epsilon\rightarrow 0} [\Psi'_{1}(x_{2}+\epsilon)-\Psi'_{1}(x_{2}-\epsilon)]=0.
\label{CD-1}
\end{align}
By regarding $\hat{z}_{}^{T}= \{c^{}_{0,1},c^{}_{0,2},c^{}_{0,3},c^{}_{0,4},c^{}_{1,1},c^{}_{1,4},c^{}_{2,2},c^{}_{2,3}\}$ as the variable vector, Eq. (\ref{CD-1}) can be written in a matrix form
\begin{align}
\mathbf{M} \cdot \hat{z}=0,
\label{MAE}
\end{align}
where
\begin{footnotesize}
\begin{align}
\mathbf{M} =\begin{pmatrix}
\psi^{}_{0,1,1}(x_{1})&\psi^{}_{0,2,1}(x_{1})&\psi^{}_{0,3,1}(x_{1})&\psi^{}_{0,4,1}(x_{1})&-\psi_{1,1,1}(x_{1})&-\psi_{1,4,1}(x_{1})&0&0\\
\psi^{}_{0,1,2}(x_{1})&\psi^{}_{0,2,2}(x_{1})&\psi^{}_{0,3,2}(x_{1})&\psi^{}_{0,4,2}(x_{1})&-\psi_{1,1,2}(x_{1})&-\psi_{1,4,2}(x_{1})&0&0\\
\psi^{}_{0,1,1}(x_{2})&\psi^{}_{0,2,1}(x_{2})&\psi^{}_{0,3,1}(x_{2})&\psi^{}_{0,4,1}(x_{2})&0&0&-\psi_{2,2,1}(x_{2})&-\psi_{2,3,1}(x_{2})\\
\psi^{}_{0,1,2}(x_{2})&\psi^{}_{0,2,2}(x_{2})&\psi^{}_{0,3,2}(x_{2})&\psi^{}_{0,4,2}(x_{2})&0&0&-\psi_{2,2,2}(x_{2})&-\psi_{2,3,2}(x_{2})\\
\gamma^{}_{0,1}\psi^{}_{0,1,1}(x_{1})&\gamma^{}_{0,2}\psi^{}_{0,2,1}(x_{1})&\gamma^{}_{0,3}\psi^{}_{0,3,1}(x_{1})&\gamma^{}_{0,4}\psi^{}_{0,4,1}(x_{1})&-\gamma^{}_{1,1}\psi_{1,1,1}(x_{1})&-\gamma^{}_{1,4}\psi_{1,4,1}(x_{1})&0&0\\
\gamma^{}_{0,1}\psi^{}_{0,1,2}(x_{1})&\gamma^{}_{0,2}\psi^{}_{0,2,2}(x_{1})&\gamma^{}_{0,3}\psi^{}_{0,3,2}(x_{1})&\gamma^{}_{0,4}\psi^{}_{0,4,2}(x_{1})&-\gamma^{}_{1,1}\psi_{1,1,2}(x_{1})&-\gamma^{}_{1,4}\psi_{1,4,2}(x_{1})&0&0\\
\gamma^{}_{0,1}\psi^{}_{0,1,1}(x_{2})&\gamma^{}_{0,2}\psi^{}_{0,2,1}(x_{2})&\gamma^{}_{0,3}\psi^{}_{0,3,1}(x_{2})&\gamma^{}_{0,4}\psi^{}_{0,4,1}(x_{2})&0&0&-\gamma^{}_{2,2}\psi_{2,2,1}(x_{2})&-\gamma^{}_{2,3}\psi_{2,3,1}(x_{2})\\
\gamma^{}_{0,1}\psi^{}_{0,1,2}(x_{2})&\gamma^{}_{0,2}\psi^{}_{0,2,2}(x_{2})&\gamma^{}_{0,3}\psi^{}_{0,3,2}(x_{2})&\gamma^{}_{0,4}\psi^{}_{0,4,2}(x_{2})&0&0&-\gamma^{}_{2,2}\psi_{2,2,2}(x_{2})&-\gamma^{}_{2,3}\psi_{2,3,2}(x_{2})
\end{pmatrix},
\label{MES}
\end{align}
\end{footnotesize}
with the two spin components $\psi^{}_{\zeta,\lambda,1}(x)=w^{}_{\zeta,\lambda}\exp(i\gamma^{}_{\zeta,\lambda} x)$ and $\psi^{}_{\zeta,\lambda,2}(x)= d^{}_{\zeta,\lambda}\exp(i\gamma^{}_{\zeta,\lambda} x)$. Obviously, the nontrivial solution to Eq.~(\ref{MES}) requires
\begin{align}
\det[ \mathbf{M}(E)]=0,
\label{DEM}
\end{align}
which yields an equation for the energy eigenvalues $E$, see Eqs.~(\ref{wk})  and (\ref{wkc}).
 Let $E_{ \Uparrow}$ and $E_{ \Downarrow}$  denote the two lowest energy solutions to Eq.~(\ref{DEM}), with $E_{ \Uparrow}\geq E_{ \Downarrow}$. The explicit expressions for the two corresponding spinors can also be completely determined by the boundary conditions   Eqs.~ (\ref{CD-1}) in conjunction with the normalization condition $\int \Psi^{\dagger}_{1,\tau=\Uparrow,\Downarrow}(x)\Psi^{}_{1,\tau}(x)dx=1$.
where $\tau=\Uparrow,~\Downarrow$ indicate the two quasi-spin eigenstates on the dot.

The two lowest-energy spin states of the other quantum dot, QD2,
i.e., $\Psi^{}_{2, \Uparrow}(x)$ and $\Psi^{}_{2,\Downarrow }(x)$,
 can be derived similarly to the discussion above. Due to the inversion symmetry  which implies  $V^{}_{\rm qd,1}(x)=V^{}_{\rm qd,2}(-x)$, the corresponding energy eigenvalues are identical to  those of QD1.

 Exploiting the localized wave functions on each quantum dot, we next derive the tunneling amplitudes of the double-quantum-dot (DQD) model, utilizing linear combinations of the atomic orbitals.
The intra-cell amplitude is expediently found by considering  the interdot tunneling in the double
 quantum-dot Hamiltonian [see Fig.~\ref{Figs1}(a)]
 \begin{align}
 H^{}_{\rm DQD-in}(x)=\frac{p^{2}}{2m_{e}}+\alpha p\sigma_{y}+\frac{\Delta^{}_{\rm z}}{2}\sigma_{\varsigma}+V^{}_{\rm DQD-in}(x)-V^{}_{\rm c},
 \label{ODQD1}
 \end{align}
with the double-well potential
\begin{align}
 V^{}_{\rm DQD-in}(x)=V^{}_{1}[1-\Theta(|x|-d)]+V^{}_{2}\Theta(|x|-L+d)
\end{align}
and the constant chemical potential $V^{}_{\rm c}=\frac{1}{2}(E^{}_{\Uparrow}+E^{}_{\Downarrow})$ to offset the on-site orbital energy (which is conventionally set at zero for a prototype SSH model~\cite{Su1979}).
To this end, one first derives the orthonormal basis $\left\{\Phi^{}_{1,\Uparrow}(x),\Phi^{}_{1,\Downarrow}(x),\Phi^{}_{2,\Uparrow}(x),
\Phi^{}_{2,\Downarrow}(x)\right\}$ using the four localized states $\Psi^{}_{1/2,\Uparrow}(x)$ and  $\Psi^{}_{1/2,\Downarrow}(x)$. Those are then used to present the electron field operators as
$
\Psi_{\rm e }(x)= \sum^{}_{\sigma=\Uparrow,\Downarrow}[a^{}_{\sigma}\Phi^{}_{1,\sigma}(x) + b^{}_{\sigma}\Phi^{}_{2,\sigma}(x)]$,
 where $a^{}_{\sigma}$ ($a^{\dagger}_{\sigma}$) and $b^{}_{\sigma}$ ($b^{\dagger}_{\sigma}$) are the annihilation (creation) operators for the quasi-spin states on QD1 and QD2,
 respectively. The second quantized form of $H^{}_{\rm DQD-in}$ is then
 \begin{align}
 H ^{}_{\rm DQD-in}=& \int \Psi^{\ast}_{\rm e}(x) H^{}_{\rm DQD-in}(x) \Psi^{}_{\rm e }(x)dx \nonumber\\
=&\frac{\Delta^{}_{\rm z}}{2} \left(a^{\dagger}_{ }\sigma^{}_{z}a^{}_{ }+b^{\dagger}_{ }\sigma^{}_{z}b^{}_{ }\right)+\left(a^{\dagger}t^{}_{\rm in}b+{\rm H.c.}\right)\ ,
\label{DQD-in}
\end{align}
with  the spinors $a^{}_{}=\{a^{}_{\Uparrow},a^{}_{\Downarrow}\}$ and $b^{}_{}= \{b^{}_{\Uparrow},b^{}_{\Downarrow}\}$. In fact, the on-site energies of different quasi-spin states in each dot are modified by the spin-orbit interaction. However, as compared to the large spin-orbit length of the InAs nanowire, i.e., $x^{}_{\rm so}=180$~nm, the small characteristic length of the dots,  i.e., $L-2d=40$~nm, renders the spin-orbit interaction induced modification negligible, and then it is a good approximation to write  the  quasi-spin up (down) on-site energy as the positive (negative) half Zeeman splitting energy as in Eq.~(\ref{DQD-in}). The quasi-spin tunneling amplitudes can be written as a matrix
\begin{align}
t^{}_{\rm in}=t^{}_{{\rm in}, 0}\exp\left[i\varphi^{}_{{\rm in} } (\hat{\mathbf{v}}^{}_{\rm in}\cdot\hat{\boldsymbol{\sigma}})\right]\ ,
\label{T-IN}
\end{align}
with
\begin{align}
[t^{}_{\rm in}]^{}_{\sigma \sigma'}=\int \Phi^{\dagger} _{1,\sigma}(x)H^{}_{\rm DQD-in}\Phi_{2,\sigma' }(x)dx\ .
\label{s19}
\end{align}

As seen, the spin-orbit interaction appears as a (matrix) phase factor multiplying the intra-cell spin tunneling matrix.
 Here, $t^{}_{\rm in,0}$ is the real tunneling amplitude dominated by the potential difference $\Delta V=V_{1}-V_{2}$ as shown in Fig.~\ref{Figs1}(c), $\varphi^{}_{\rm in }$ is the spin rotation phase induced by the spin-orbit interaction, and $\hat{\mathbf{v}}^{}_{\rm in}=\{\sin(\vartheta^{}_{\rm in}),0,\cos(\vartheta^{}_{\rm in})\}$ is the unit vector determining the direction of the axis of rotation. Numerically, $\varphi^{}_{\rm in }$ is proportional to the ratio between the interdot distance and the spin-orbit length $L/x^{}_{\rm so}$ with $x^{}_{\rm so}= \hbar/(m^{}_{e}\alpha)$,  and that is consistent with
the analytical analysis in Ref.~\onlinecite{Liu2018}.

It is important to note that the coordinate system for defining the rotation vector $\hat{\mathbf{v}}^{}_{{\rm in}}$
 does not coincide with that defining the spin operators in Eq.~(\ref{ODQD1}).
The $x$-component of $\hat{\mathbf{v}}^{}_{{\rm in}}$ corresponds to the interdot spin-flip tunneling between the localized quasi-spin states, and vanishes in the absence of spin mixing.
Therefore, for zero magnetic field, i.e., when $\Delta^{}_{\rm z}=0$, or when $\theta=0$, the vector $\hat{\mathbf{v}}^{}_{\rm in}$ is purely along the $z$ direction since then $[H^{}_{\rm DQD-in},\sigma^{}_{y}]=0$. In other cases,
the specific value of $\vartheta^{}_{\rm in }$ depends on the degree of spin mixing on each localized quasi-spin state, which in turn is determined by the angle between the external magnetic field and the one induced by the spin-orbit  interaction. The amplitude of the spin-flip tunneling in Eq.~(\ref{T-IN}), $t^{\prime}_{\rm in }=t^{}_{\rm in, 0}\sin(\varphi^{}_{\rm in }) \sin(\vartheta^{}_{\rm in })$, as a function of the tilting angle $\theta$ for a fixed magnitude of the magnetic field, is portrayed in Fig.~\ref{Figs1}(d), showing that it vanishes for $\theta=0$.

The inter-cell hopping is addressed by considering a neighboring DQD [as shown in Fig.~\ref{Figs1}(b)] which obeys the Hamiltonian
\begin{align}
H^{}_{\rm DQD-ex}(x)=\frac{p^{2}}{2m_{e}}+\alpha p \sigma_{y}+\frac{\Delta_{\rm z}}{2}\sigma^{}_{\varsigma}+V^{}_{\rm DQD-ex}(x)-V^{}_{\rm c},
\end{align}
where the double-well potential takes the form
\begin{align}
V^{}_{\rm DQD-ex}(x)=V^{}_{2}[1-\Theta(|x|-d)]+V^{}_{1}\Theta(|x|-L+d).
\end{align}
Similar to the intra-cell spin tunneling discussed above, the spin transfer matrix for the inter-cell tunneling  takes the form
\begin{align}
t^{}_{\rm ex}= t^{}_{{\rm ex},0}\exp\left[i\varphi^{}_{{\rm ex} } (\hat{\mathbf{v}}^{}_{\rm ex}\cdot\hat{\boldsymbol{\sigma}})\right]\ ,
\label{T-EX}
\end{align}
with $t^{}_{{\rm ex},0}$ and $\varphi^{}_{{\rm ex}}$ corresponding to the real amplitude and the spin rotation phase of the interdot tunneling, and $\hat{\mathbf{v}}^{}_{{\rm ex}}=\{\sin(\vartheta^{}_{{\rm ex}}),0,\cos(\vartheta^{}_{{\rm ex}})\}$  describing the direction of the axis of rotation.
 Analogously, the corresponding spin-flip tunneling amplitude  $t^{\prime}_{\rm ex}=t^{}_{{\rm ex},0}\sin(\varphi^{}_{\rm ex})\sin(\vartheta^{}_{\rm ex})$ can be regulated by manipulating the tilting angle $\theta$, as shown in Fig.~\ref{Figs1}(d).
Introducing the intra-cell spin tunneling, the second quantized Hamiltonian of the double-quantum-dot chain is
\begin{align}
H^{}_{\rm T}= \sum^{}_{n}\frac{\Delta^{}_{\rm z}}{2}\left(a^{\dagger}_{n} \sigma^{}_{z}a^{}_{n}+b^{\dagger}_{n}\sigma^{}_{z}b_{n} \right)+\sum^{}_{n}\left(a^{\dagger}_{n}t^{}_{\rm in }b^{}_{n}+a^{\dagger}_{n+1}t^{}_{\rm ex}b^{}_{n}+{\rm h.c.}\right)\ ,
\label{TBT}
\end{align}
where $n$ is the index of the unit cell,  and   $a^{}_{n}=\{a^{}_{n\Uparrow},a^{}_{n\Downarrow}\}$ and $b^{}_{n}=\{b^{}_{n\Uparrow},b^{}_{n\Downarrow}\}$  are spinors.

For a spinless DQD chain, the tunneling matrices in Eqs.~(\ref{T-IN}) and (\ref{T-EX}) become real numbers, $t^{}_{{\rm in},0}$ and $t^{}_{{\rm ex},0}$, and the effective tight-binding Hamiltonian  is identical to that of the single-level Su-Schrieffer-Heeger model~\cite{Su1979}
\begin{align}
H^{}_{{\rm T},0}=\sum^{}_{n}\left(t^{}_{{\rm in},0 } a^{\dagger}_{n}b^{}_{n}+t^{}_{{\rm ex},0 } a^{\dagger}_{n+1}b^{}_{n}+{\rm h.c.}\right),
\label{TB0}
\end{align}
with $a^{}_{n}$ and $b^{}_{n}$ corresponding to the operators  for the lowest localized orbitals on the QD1 and QD2, respectively.


\section{The Bloch spectrum of the double-quantum-dot chain ~\label{S-2}}

\begin{figure}
\centering
\includegraphics[width=0.4\textwidth]{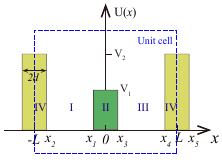}
\caption{ (color online) The spatial distribution of the confinement potential $U(x)$ in a unit cell; the separation of the unit cell into four regions (denoted by $I$, $II$, $III$ and $IV$)  in which the Bloch functions are derived (see text).}
\label{Figs2}
\end{figure}

Here the detailed calculation of the Bloch bands of our system is given.
For a one-dimensional lattice obeying periodic boundary conditions, the eigenstates are  the Bloch wave functions
\begin{align}
\Psi^{}_{k}(x)=\exp\left(ikx\right)u^{}_{k}(x),
\end{align}
 where $k$ is the wave vector and $u^{}_{k}(x)$ is the cell-periodic Bloch wave function. Obviously, the corresponding energy spectrum $\varepsilon(k)$ is given by
 \begin{align}
 H \Psi^{}_{k}(x)=\varepsilon(k)\Psi^{}_{k}(x),
 \label{B-F}
 \end{align}
where
 \begin{align}
  H =\frac{p^{2}}{2m_{e}}-V^{}_{\rm c}+U(x)+\alpha p \sigma_{y}+\frac{\Delta^{}_{\rm z}}{2}\sigma^{}_{\varsigma}.
\end{align}

 Similarly to the derivation in Appendix \ref{S-1}, the solutions of Eq.~(\ref{B-F}) are obtained from the eigenfunctions of the Schr\"{o}dinger equations with different gate potentials
\begin{align}
\widetilde{H}^{}_{\zeta=0,1,2}\Psi^{}_{k,\zeta }(x)=E \Psi_{k,\zeta}(x).
\label{BSZ}
\end{align}
where $\Psi^{}_{k,\zeta}(x)=\exp\left(ikx\right)u^{}_{k,\zeta}(x)$  and $\varepsilon(k)=E-V^{}_{\rm c}$ (measured from the chemical potential $V_{\rm c}$).
 Then, analogously to the eigenstates in Eq.~(\ref{LQD}),  there exist four-fold degenerate solutions to Eq.~(\ref{BSZ}) for a fixed $k$, of the form
 \begin{align}
u^{}_{k,\zeta,\lambda}(x)=\exp\big[i(\gamma^{}_{\zeta,\lambda}-k)x\big]\begin{pmatrix}
w^{}_{\zeta,\lambda}\\
d^{}_{\zeta,\lambda}
\end{pmatrix}
\label{BF}
\end{align}
with the wave vector $\gamma^{}_{\zeta,\lambda}$ and the two spin components ($u^{}_{\zeta,\lambda}$ and $d^{}_{\zeta,\lambda}$) determined by Eqs.~(\ref{wk}) and (\ref{wkc}). It is convenient to divide the unit cell into four parts,  as shown in Fig.~\ref{Figs2}. The Bloch functions in each segment are expanded as
\begin{align}
u^{}_{k,I} (x)=&\sum^{4}_{\lambda=1}c^{}_{I,\lambda}u^{}_{k,0,\lambda} (x)~~~~ u^{}_{k,II} (x)=\sum^{4}_{\lambda=1}c^{}_{II,\lambda}u^{}_{k,1,\lambda}(x) \nonumber\\
u^{}_{k,III} (x)=&\sum^{4}_{\lambda=1}c^{}_{III,\lambda}u^{}_{k,0,\lambda}(x) ~~~~ u^{}_{k,IV} (x)=\sum^{4}_{\lambda=1}c^{}_{IV,\lambda}u^{}_{k,2,\lambda}(x),
\label{4u}
\end{align}
with $c^{}_{\Lambda,\lambda}$ $(\Lambda=I,~II,~III,~IV)$ representing the, yet to be determined, sixteen coefficients. The entire Bloch wave function is then
\begin{align}
u^{}_{k}(x)=&\Theta(x+L-d)[1-\Theta(x+d)]u^{}_{k,I}(x)+[1-\Theta(|x|-d)]u^{}_{k,II}(x)+[1-\Theta(x-L+d)]\nonumber\\
&\times  \Theta(x-d) u^{}_{k,III}(x) + \Theta(|x|-L+d)[1-\Theta(|x|-L)]u^{}_{k,IV}(x).
\label{bfc}
\end{align}
Combined with the periodicity of the cell Bloch functions, i.e., $u^{}_{k}(x_{2})=u^{}_{k}(x_{5})$ and $u^{\prime}_{k}(x_{2})=u^{\prime}_{k}(x_{5})$ with $x^{}_{5}\equiv x^{}_{2}+2L= L+d$, the energy bands are determined by the continuity of the wave function at the interfaces, $\lim_{\epsilon \rightarrow 0 }[u^{}_{k}(\xi+\epsilon)-u^{}_{k}(\xi+\epsilon)]=0$ and $\lim_{\epsilon \rightarrow 0 }[u^{\prime}_{k}(\xi+\epsilon)-u^{\prime}_{k}(\xi+\epsilon)]=0$,  with $\xi=x^{}_{1},x^{}_{3},x^{}_{4}$. Regarding the coefficients of the combinations in Eqs. (\ref{4u}) as a variable vector, the prerequisite conditions can be written in a matrix form
\begin{align}
\mathbf{W}(k,E) \cdot \hat{s}=0,
\label{MAT2}
\end{align}
where the vector is
$\hat{s}^{T}=\{c^{}_{I,1},c^{}_{I,2},c^{}_{I,3},c^{}_{I,4},c^{}_{II,1},c^{}_{II,2},c^{}_{II,3},c^{}_{II,4},c^{}_{III,1},c^{}_{III,2},c^{}_{III,3},c^{}_{III,4},c^{}_{IV,1},c^{}_{IV,2},c^{}_{IV,3},c^{}_{IV,4}\}$ and the matrix is
\begin{footnotesize}
\begin{align}
\mathbf{W}=\left(\begin{array}{cccccccccccccccc}
u^{(1,\uparrow)}_{k,0,\lambda}&u^{(1,\uparrow)}_{k,0,\lambda}&u^{(1,\uparrow)}_{k,0,\lambda}&u^{(1,\uparrow)}_{k,0,\lambda}&-u^{(1,\uparrow)}_{k,1,\lambda}&-u^{(1,\uparrow)}_{k,1,\lambda}&-u^{(1,\uparrow)}_{k,1,\lambda} &-u^{(1,\uparrow)}_{k,1,\lambda} &0&0&0&0 &0&0 &0&0\\
u^{(1,\downarrow)}_{k,0,\lambda}&u^{(1,\downarrow)}_{k,0,\lambda}&u^{(1,\downarrow)}_{k,0,\lambda}&u^{(1,\downarrow)}_{k,0,\lambda}&-u^{(1,\downarrow)}_{k,1,\lambda}&-u^{(1,\downarrow)}_{k,1,\lambda}&-u^{(1,\downarrow)}_{k,1,\lambda} &-u^{(1,\downarrow)}_{k,1,\lambda} &0&0&0&0 &0&0 &0&0\\
0&0&0&0&-u^{(3,\uparrow)}_{k,1,\lambda}&-u^{(3,\uparrow)}_{k,1,\lambda}&-u^{(3,\uparrow)}_{k,1,\lambda} &-u^{(3,\uparrow)}_{k,1,\lambda} &u^{(3,\uparrow)}_{k,0,\lambda}&u^{(3,\uparrow)}_{k,0,\lambda}&u^{(3,\uparrow)}_{k,0,\lambda}&u^{(3,\uparrow)}_{k,0,\lambda} &0&0 &0&0\\
0&0&0&0&-u^{(3,\downarrow)}_{k,1,\lambda}&-u^{(3,\downarrow)}_{k,1,\lambda}&-u^{(3,\downarrow)}_{k,1,\lambda} &-u^{(3,\downarrow)}_{k,1,\lambda} &u^{(3,\downarrow)}_{k,0,\lambda}&u^{(3,\downarrow)}_{k,0,\lambda}&u^{(3,\downarrow)}_{k,0,\lambda}&u^{(3,\downarrow)}_{k,0,\lambda} &0&0 &0&0\\
0&0&0&0&0&0&0 &0&u^{(4,\uparrow)}_{k,0,\lambda}&u^{(4,\uparrow)}_{k,0,\lambda}&u^{(4,\uparrow)}_{k,0,\lambda}&u^{(4,\uparrow)}_{k,0,\lambda} &-u^{(4,\uparrow)}_{k,2,\lambda}&-u^{(4,\uparrow)}_{k,2,\lambda}&-u^{(4,\uparrow)}_{k,2,\lambda} &-u^{(4,\uparrow)}_{k,2,\lambda}\\
0&0&0&0&0&0&0 &0&u^{(4,\downarrow)}_{k,0,\lambda}&u^{(4,\downarrow)}_{k,0,\lambda}&u^{(4,\downarrow)}_{k,0,\lambda}&u^{(4,\downarrow)}_{k,0,\lambda} &-u^{(4,\downarrow)}_{k,2,\lambda}&-u^{(4,\downarrow)}_{k,2,\lambda}&-u^{(4,\downarrow)}_{k,2,\lambda} &-u^{(4,\downarrow)}_{k,2,\lambda}\\
u^{(2,\uparrow)}_{k,0,\lambda}&u^{(2,\uparrow)}_{k,0,\lambda}&u^{(2,\uparrow)}_{k,0,\lambda}&u^{(2,\uparrow)}_{k,0,\lambda}&0&0&0 &0 &0&0&0&0 &-u^{(5,\uparrow)}_{k,2,\lambda}&-u^{(5,\uparrow)}_{k,2,\lambda}&-u^{(5,\uparrow)}_{k,2,\lambda} &-u^{(5,\uparrow)}_{k,2,\lambda}\\
u^{(2,\downarrow)}_{k,0,\lambda}&u^{(2,\downarrow)}_{k,0,\lambda}&u^{(2,\downarrow)}_{k,0,\lambda}&u^{(2,\downarrow)}_{k,0,\lambda}&0&0&0 &0 &0&0&0&0 &-u^{(5,\downarrow)}_{k,2,\lambda}&-u^{(5,\downarrow)}_{k,2,\lambda}&-u^{(5,\downarrow)}_{k,2,\lambda} &-u^{(5,\downarrow)}_{k,2,\lambda}\\
\tilde{u}^{(1,\uparrow)}_{k,0,\lambda}&\tilde{u}^{(1,\uparrow)}_{k,0,\lambda}&\tilde{u}^{(1,\uparrow)}_{k,0,\lambda}&\tilde{u}^{(1,\uparrow)}_{k,0,\lambda}&-\tilde{u}^{(1,\uparrow)}_{k,1,\lambda}&-\tilde{u}^{(1,\uparrow)}_{k,1,\lambda}&-\tilde{u}^{(1,\uparrow)}_{k,1,\lambda} &-\tilde{u}^{(1,\uparrow)}_{k,1,\lambda} &0&0&0&0 &0&0 &0&0\\
\tilde{u}^{(1,\downarrow)}_{k,0,\lambda}&\tilde{u}^{(1,\downarrow)}_{k,0,\lambda}&\tilde{u}^{(1,\downarrow)}_{k,0,\lambda}&\tilde{u}^{(1,\downarrow)}_{k,0,\lambda}&-\tilde{u}^{(1,\downarrow)}_{k,1,\lambda}&-\tilde{u}^{(1,\downarrow)}_{k,1,\lambda}&-\tilde{u}^{(1,\downarrow)}_{k,1,\lambda} &-\tilde{u}^{(1,\downarrow)}_{k,1,\lambda} &0&0&0&0 &0&0 &0&0\\
0&0&0&0&-\tilde{u}^{(3,\uparrow)}_{k,1,\lambda}&-\tilde{u}^{(3,\uparrow)}_{k,1,\lambda}&-\tilde{u}^{(3,\uparrow)}_{k,1,\lambda} &-\tilde{u}^{(3,\uparrow)}_{k,1,\lambda} &\tilde{u}^{(3,\uparrow)}_{k,0,\lambda}&\tilde{u}^{(3,\uparrow)}_{k,0,\lambda}&\tilde{u}^{(3,\uparrow)}_{k,0,\lambda}&\tilde{u}^{(3,\uparrow)}_{k,0,\lambda} &0&0 &0&0\\
0&0&0&0&-\tilde{u}^{(3,\downarrow)}_{k,1,\lambda}&-\tilde{u}^{(3,\downarrow)}_{k,1,\lambda}&-\tilde{u}^{(3,\downarrow)}_{k,1,\lambda} &-\tilde{u}^{(3,\downarrow)}_{k,1,\lambda} &\tilde{u}^{(3,\downarrow)}_{k,0,\lambda}&\tilde{u}^{(3,\downarrow)}_{k,0,\lambda}&\tilde{u}^{(3,\downarrow)}_{k,0,\lambda}&\tilde{u}^{(3,\downarrow)}_{k,0,\lambda} &0&0 &0&0\\
0&0&0&0&0&0&0 &0&\tilde{u}^{(4,\uparrow)}_{k,0,\lambda}&\tilde{u}^{(4,\uparrow)}_{k,0,\lambda}&\tilde{u}^{(4,\uparrow)}_{k,0,\lambda}&\tilde{u}^{(4,\uparrow)}_{k,0,\lambda} &-\tilde{u}^{(4,\uparrow)}_{k,2,\lambda}&-\tilde{u}^{(4,\uparrow)}_{k,2,\lambda}&-\tilde{u}^{(4,\uparrow)}_{k,2,\lambda} &-\tilde{u}^{(4,\uparrow)}_{k,2,\lambda}\\
0&0&0&0&0&0&0 &0&\tilde{u}^{(4,\downarrow)}_{k,0,\lambda}&\tilde{u}^{(4,\downarrow)}_{k,0,\lambda}&\tilde{u}^{(4,\downarrow)}_{k,0,\lambda}&\tilde{u}^{(4,\downarrow)}_{k,0,\lambda} &-\tilde{u}^{(4,\downarrow)}_{k,2,\lambda}&-\tilde{u}^{(4,\downarrow)}_{k,2,\lambda}&-\tilde{u}^{(4,\downarrow)}_{k,2,\lambda} &-\tilde{u}^{(4,\downarrow)}_{k,2,\lambda}\\
\tilde{u}^{(2,\uparrow)}_{k,0,\lambda}&\tilde{u}^{(2,\uparrow)}_{k,0,\lambda}&\tilde{u}^{(2,\uparrow)}_{k,0,\lambda}&\tilde{u}^{(2,\uparrow)}_{k,0,\lambda}&0&0&0 &0 &0&0&0&0 &-\tilde{u}^{(5,\uparrow)}_{k,2,\lambda}&-\tilde{u}^{(5,\uparrow)}_{k,2,\lambda}&-\tilde{u}^{(5,\uparrow)}_{k,2,\lambda} &-\tilde{u}^{(5,\uparrow)}_{k,2,\lambda}\\
\tilde{u}^{(2,\downarrow)}_{k,0,\lambda}&\tilde{u}^{(2,\downarrow)}_{k,0,\lambda}&\tilde{u}^{(2,\downarrow)}_{k,0,\lambda}&\tilde{u}^{(2,\downarrow)}_{k,0,\lambda}&0&0&0 &0 &0&0&0&0 &-\tilde{u}^{(5,\downarrow)}_{k,2,\lambda}&-\tilde{u}^{(5,\downarrow)}_{k,2,\lambda}&-\tilde{u}^{(5,\downarrow)}_{k,2,\lambda} &-\tilde{u}^{(5,\downarrow)}_{k,2,\lambda}
\end{array}\right)
\label{CMAT}
\end{align}
\end{footnotesize}
with $u^{(j,\uparrow)}_{k,\zeta,\lambda}= \exp[i(\gamma^{}_{\zeta,\lambda}-k)x_{j}]w^{}_{\zeta,\lambda}$ ($j=1,2,3,4,5$), $u^{(j,\downarrow)}_{k,\zeta,\lambda}= \exp[i(\gamma^{}_{\zeta,\lambda}-k)x_{j}]d^{}_{\zeta,\lambda}$, $\tilde{u}^{(j,\uparrow)}_{k,\zeta,\lambda}= (\gamma^{}_{\zeta,\lambda}-k)\exp[i(\gamma^{}_{\zeta,\lambda}-k)x_{j}]w^{}_{\zeta,\lambda}$, and $\tilde{u}^{(j,\downarrow)}_{k,\zeta,\lambda}= (\gamma^{}_{\zeta,\lambda}-k)\exp[i(\gamma^{}_{\zeta,\lambda}-k)x_{j}]d^{}_{\zeta,\lambda}$. A nontrivial solution to Eq.~(\ref{MAT2}) requires
\begin{align}
\det [\mathbf{W} (k,E)]=0,
\label{GF}
\end{align}
 and yields an implicit equation for $k$ and $E$ of Eqs.~(\ref{wk}),~(\ref{wkc}), and ~(\ref{CMAT}). The exact energy spectrum of the Bloch band $\varepsilon (k)=E-V^{}_{\rm c}$ is found by solving numerically for $E$ at a fixed value of $k$. Substituting the so-derived values of $E$ in Eq.~(\ref{MAT2}) yields the values of the coefficients $c^{}_{\Lambda,\lambda}$, i.e., gives the specific form of the Bloch wave functions, with the normalization condition $\int^{L}_{-L}u^{\ast}_{k}(x)u^{ }_{k}(x)dx=1$.

In the absence of a magnetic field or when $\theta=0$, the two spin degrees of freedom are separable and the number of the corresponding coefficients  for ascertaining the Bloch band is reduced by half, which is similar to a spinless chain.  In this case, the matrix $\mathbf{W}(k,E)$ is  decomposed into two independent submatrices whose determinants yield the Bloch spectrum, as shown in Fig.~\ref{Fig2}(c) for the case of   zero magnetic field.


%

\end{widetext}
\end{appendix}



\begin{thebibliography}{99}

 \bibitem{Moore2010} J. E. Moore,  The birth of topological insulators, \href{https://www.nature.com/articles/nature08916}{Nature {\bf464}, 194 (2010)}.
\bibitem{Hasan2010}M. Z. Hasan and C. L. Kane, Colloquium:Topological insulators, \href{https://link.aps.org/doi/10.1103/RevModPhys.82.3045}{Rev. Mod. Phys. {\bf82}, 3045  (2010)}.
\bibitem{Ryu}
S. Ryu and Y. Hatsugai, Topological Origin of Zero-Energy Edge States in Particle-Hole Symmetric Systems, \href{https://journals.aps.org/prl/abstract/10.1103/PhysRevLett.89.077002}{\prl{89}, 077002 (2002)}; T. Kawarabayashi and Y. Hatsugai, Bulk-edge correspondence
 with generalized chiral symmetry, \href{https://journals.aps.org/prb/pdf/10.1103/PhysRevB.103.205306}{\prb{103}, 205306 (2021)}.
\bibitem{Su1979}W. P. Su, J. R. Schrieffer, and A. J. Heeger,  Solitons in Polyacetylene, \href{https://journals.aps.org/prl/abstract/10.1103/PhysRevLett.42.1698}{\prl{42}, 1698 (1979)}.
\bibitem{Rice1982}M. J. Rice and E. J. Mele, Elementary Excitations of a Linearly Conjugated Diatomic Polymer, \href{https://journals.aps.org/prl/abstract/10.1103/PhysRevLett.49.1455}{\prl{49}, 1455 (1982)}.

 \bibitem{Li2014} L. Li, Z. Xu, and S. Chen, Topological phases of generalized Su-Schrieffer-Heeger models, \href{https://journals.aps.org/prb/abstract/10.1103/PhysRevB.89.085111}{\prb{89}, 085111 (2014)}.
\bibitem{Xie2019}D. Xie, W. Gou, T. Xiao, B. Gadway, and B. Yan, Topological characterizations of an extended
Su-Schrieffer-Heeger model, \href{https://doi.org/10.1038/s41534-019-0159-6}{npj Quantum Information {\bf5}, 55 (2019)}.
\bibitem{Zhu2014}B. Zhu, R. Lu, and S. Chen, PT symmetry in the non-Hermitian Su-Schrieffer-Heeger model with complex boundary potentials, \href{https://journals.aps.org/pra/abstract/10.1103/PhysRevA.89.062102}{\pra{89}, 062102 (2014)}.
\bibitem{Lieu2018}S. Lieu,  Topological phases in the non-Hermitian Su-Schrieffer-Heeger model, \href{https://journals.aps.org/prb/abstract/10.1103/PhysRevB.97.045106}{\prb{97}, 045106 (2018)}.
\bibitem{Kunst2018}F. K. Kunst,  E. Edvardsson, J. C. Budich, and E. J. Bergholtz,  Biorthogonal Bulk-Boundary Correspondence in Non-Hermitian Systems, \href{https://journals.aps.org/prl/abstract/10.1103/PhysRevLett.121.026808}{\prl{121}, 026808 (2018)}.
\bibitem{Yao2018}S. Yao and Z. Wang,  Edge States and Topological Invariants of Non-Hermitian Systems, \href{https://journals.aps.org/prl/abstract/10.1103/PhysRevLett.121.086803}{\prl{121}, 086803 (2018)}.
\bibitem{Chen2019}R. Chen, C.-Z. Chen,  B. Zhou,  and D.-H. Xu,  Finite-size effects in non-Hermitian topological systems, \href{https://journals.aps.org/prb/abstract/10.1103/PhysRevB.99.155431}{\prb{99}, 155431 (2019)}.
 \bibitem{Lang2012}L.-J. Lang, X. Cai, and S. Chen, Edge states and Topological phases in one-dimensional optical superlattices, \href{https://journals.aps.org/prl/abstract/10.1103/PhysRevLett.108.220401}{\prl{108}, 220401 (2012)}.
\bibitem{Atala2013} M. Atala, M. Aidelsburger, J. T. Barreiro, D. Abanin, T. Kitagawa, E. Demler and I. Bloch, Direct measurement of the Zak phase in topological Bloch bands,  \href{https://www.nature.com/articles/nphys2790}{\nap{9}, 795 (2013)}.
 \bibitem{Meier2016}E. J. Meier, F. Alex An, and B. Gadway, Observation of the topological soliton state in the Su-Schrieffer-Heeger model, \href{https://www.nature.com/articles/ncomms13986}{\nac{7}, 13986 (2016)}.
\bibitem{Leder2016}M. Leder, C. Grossert, L. Sitta, M. Genske, A. Rosch, and  M. Weitz, Real-space imaging of a topologically protected edge state with ultracold atoms in an amplitude-chirped optical lattice, \href{https://www.nature.com/articles/ncomms13112}{\nac{7}, 13112 (2016)}.
\bibitem{Solnyshkov2016}D. D. Solnyshkov, A. V. Nalitov, and G. Malpuech, Kibble-Zurek Mechanism in Topologically Nontrivial Zigzag Chains of Polariton Micropillars, \href{https://journals.aps.org/prl/abstract/10.1103/PhysRevLett.116.046402}{\prl{116}, 046402 (2016)}.
\bibitem{Jean2017}P. St-Jean, V. Goblot, E. Galopin, A. Lema\^{\i}tre, T. Ozawa, L. Le Gratiet, I. Sagnes, J. Bloch, and A. Amo, Lasing in topological edge states of a one-dimensional lattice, \href{https://www.nature.com/articles/s41566-017-0006-2}{\napt{11}, 651
(2017)}.
\bibitem{Parto2018}M. Parto, S. Wittek, H. Hodaei, G. Harari, M. A. Bandres, J. Ren, M. C. Rechtsman, M. Segev, D. N. Christodoulides, and M. Khajavikhan, Edge-Mode Lasing in 1D Topological Active Arrays, \href{https://journals.aps.org/prl/abstract/10.1103/PhysRevLett.120.113901}{\prl{120}, 113901 (2018)}.
\bibitem{Whittaker2019}C. E. Whittaker, E. Cancellieri, P. M. Walker,B. Royall, L. E. Tapia Rodriguez, E. Clarke, D. M. Whittaker,
H. Schomerus, M. S. Skolnick, and D. N. Krizhanovskii, Effect of photonic spin-orbit coupling on the topological edge modes of a Su-Schrieffer-Heeger chain, \href{https://journals.aps.org/prb/abstract/10.1103/PhysRevB.99.081402}{\prb{99}, 081402(R) (2019)}.
\bibitem{Groning2018}O. Gr\"{o}ning, S. Wang, X. Yao, C. A. Pignedoli, G. B. Barin, C. Daniels, A. Cupo, V. Meunier, X. Feng, A. Narita, K. M\"{u}llen, P. Ruffieux, and R. Fasel, Engineering of robust topological quantum phases in graphene nanoribbons, \href{https://www.nature.com/articles/s41586-018-0375-9}{\nat{560}, 209 (2018)}.
\bibitem{Drost2017}R. Drost, T. Ojanen, A. Harju, and P. Liljeroth,  Topological states in engineered atomic lattices, \href{https://www.nature.com/articles/nphys4080}{\nap{13}, 668 (2017)}.

\bibitem{Rizzo2018}D. J. Rizzo, G. Veber, T. Cao, C. Bronner, T. Chen, F. Zhao, H. Rodriguez, S. G. Louie, M. F. Crommie, and F. R. Fischer,  Topological band engineering of graphene nanoribbons, \href{https://www.nature.com/articles/s41586-018-0376-8}{\nat{560}, 204 (2018)}.
\bibitem{Kane2005}C. L. Kane and E. J. Mele, Z$_
{2}$ Topological Order and the Quantum Spin Hall Effect, \href{https://link.aps.org/doi/10.1103/PhysRevLett.95.146802}{Phys. Rev. Lett. {\bf95}, 146802 (2005)}; Quantum Spin Hall Effect in Graphene, \href{https://link.aps.org/doi/10.1103/PhysRevLett.95.226801}{Phys. Rev. Lett. {\bf95}, 226801 (2005)}.
\bibitem{Sau2010}J. D. Sau, R. M. Lutchyn, S. Tewari, and S. Das Sarma, Generic New Platform for Topological Quantum Computation
Using Semiconductor Heterostructures, \href{https://journals.aps.org/prl/abstract/10.1103/PhysRevLett.104.040502}{\prl{104}, 040502 (2010)}.
\bibitem{Yan2014}Z. Yan and S. Wan, Topological phases, topological flat bands, and topological excitations in a one-dimensional dimerized lattice with spin-orbit coupling, \href{doi: 10.1209/0295-5075/107/47007}{Europhys. Lett.{\bf107}, 47007 (2014)}.
\bibitem{Bahari2016}M. Bahari and M. V. Hosseini, Zeeman-field-induced nontrivial topological phases in a one-dimensional
spin-orbit-coupled dimerized lattice, \href{https://journals.aps.org/prb/abstract/10.1103/PhysRevB.94.125119}{\prb{94}, 125119 (2016)}.
\bibitem{AC1984} Y. Aharonov and A. Casher,
 Topological Quantum Effects for Neutral Particles,
 \href{https://link.aps.org/doi/10.1103/PhysRevLett.53.319}{Phys. Rev. Lett. {\bf 53}, 319 (1984)}.
\bibitem{Yao2017}Y. Yao, M. Sato, T. Nakamura, N. Furukawa, and M. Oshikawa, Theory of electron spin resonance in one-dimensional topological insulators with spin-orbit couplings: Detection of edge states, \href{https://journals.aps.org/prb/abstract/10.1103/PhysRevB.96.205424}{\prb{96}, 205424 (2017)}.
\bibitem{Shahbazyan1994}
T. V. Shahbazyan and M. E. Raikh,
 Low-Field Anomaly in 2D Hopping Magnetoresistance Caused by Spin-Orbit Term in the Energy Spectrum,
 \href{https://link.aps.org/doi/10.1103/PhysRevLett.73.1408}{Phys. Rev. Lett. {\bf 73}, 1408 (1994)};
O. Entin-Wohlman and A. Aharony,
 DC Spin geometric phases in hopping magnetoconductance,
\href{https://link.aps.org/doi/10.1103/PhysRevResearch.1.033112}{Phys. Rev. Research {\bf 1}, 033112 (2019)}.

\bibitem{Zhang2017} W.-F. Zhang, C.-Y. Li, X.-F. Chen, C.-M. Huang, F.-W. Ye, Topological zero-energy modes in time-reversal-symmetry-broken systems, \href{http://wulixb.iphy.ac.cn/en/article/Y2017/V66/I22/220201}{Acta Physica Sinica {\bf66}, 220201 (2017)}.
\bibitem{Grusdt2013}F. Grusdt, M. H\"{o}ning, and M. Fleischhauer, Topological Edge States in the One-Dimensional Superlattice Bose-Hubbard Model, \href{https://journals.aps.org/prl/abstract/10.1103/PhysRevLett.110.260405}{\prl{110}, 260405 (2013)}.
\bibitem{Liu2018b}Z.-H. Liu, R. Li, X. Hu, and J. Q. You, Spin-orbit coupling and electric-dipole spin resonance in a nanowire double quantum dot, \href{https://www.nature.com/articles/s41598-018-20706-5}{Sci. Rep. {\bf8}, 2302 (2018)}.
\bibitem{Winkler2003} R. Winkler, {\it Spin-orbit Coupling Effects in Two-dimensional Electron and Hole Systems}, (Springer,
Berlin, 2003).
\bibitem{Scherubl2016}Z. Scher\"{u}bl, G. F\"{u}l\"{o}p, M. H. Madsen, J. Nyg{\aa}rd, and S. Csonka, Electrical tuning of Rashba spin-orbit interaction in multigated InAs nanowires, \href{https://journals.aps.org/prb/cited-by/10.1103/PhysRevB.94.035444}{\prb{94}, 035444 (2016)}.


 \bibitem{Supplement0}The sum of the two periodic barrier potentials is fixed as 40~meV to ensure effective coupling between two adjacent dots when manipulating the dimerization strength, scaled by the potential difference $\Delta V=V^{}_{1}-V^{}_{2}$.



\bibitem{Liu2018}Z.-H. Liu, O. Entin-Wohlman, A. Aharony, and J. Q. You,  Control of the two-electron exchange interaction in a nanowire double quantum dot, \href{https://journals.aps.org/prb/abstract/10.1103/PhysRevB.98.241303}{\prb{98}, 241303(R) (2018)}.
\bibitem{Delplace2011}
P. Delplace, D. Ullmo, and G. Montambaux, Zak phase and the existence of edge states in graphene, \href{https://journals.aps.org/prb/abstract/10.1103/PhysRevB.84.195452}{\prb{84}, 195452 (2011)}.
\bibitem{Supplement2}The consistency between the  continuous Hamiltonian [see Eq.~(\ref{H-0})] and the discrete model in Eq.~(\ref{T-B}) requires  the conditions   $L\gg4d$   and $\min\{V^{}_{1},V^{}_{2}\}\gg \max\{\Delta^{}_{\rm z}, m^{}_{e}\alpha^{2} \} $.


\bibitem{Zak1989}J. Zak, Berry's phase for energy bands in solids, \href{https://journals.aps.org/prl/abstract/10.1103/PhysRevLett.62.2747}{\prl{62}, 2747 (1989)}.
\bibitem{Hatsugai2006} Y. Hatsugai, Quantized Berry phases as a local order parameter of a Quantum Liquid, \href{https://doi.org/10.1143/JPSJ.75.123601}{J. Phys. Soc. Jpn. {\bf75}, 123601 (2006)}.
\bibitem{Fukui2005}T. Fukui, Y. Hatsugai, and H. Suzuki, Chern Numbers in Discretized Brillouin Zone:
Efficient Method of Computing (Spin) Hall Conductances, \href{https://journals.jps.jp/doi/10.1143/JPSJ.74.1674}{J. Phys. Soc. Jpn. {\bf 74}, 1674 (2005)}.
\bibitem{Asboth2016}J. K. Asb\'{o}the, L. Oroszl\'{a}ny, and A. P\'{a}lyi,
The Su-Schrieffer-Heeger (SSH) Model,  \href{https://link.springer.com/chapter/10.1007/978-3-319-25607-8_1}{ Lect. Notes Phys. {\bf919}, 1 (2016)}.
\bibitem{Nevado2017}P. Nevado, S. Fern\'{a}ndez-Lorenzo, and D. Porras, Topological Edge States in Periodically Driven Trapped-Ion Chains, \href{https://journals.aps.org/prl/pdf/10.1103/PhysRevLett.119.210401}{\prl{119}, 210401 (2017)}.
\bibitem{Li2013}R. Li,  J. Q. You, C. P. Sun, and F. Nori, Controlling a Nanowire Spin-Orbit Qubit via Electric-Dipole Spin Resonance,  \href{https://journals.aps.org/prl/abstract/10.1103/PhysRevLett.111.086805}{\prl{111}, 086805 (2013)}.
\bibitem{Nilsson2009}H. A. Nilsson, P. Caroff, C. Thelander, M. Larsson, J. B. Wagner, L.-E. Wernersson, L. Samuelson, and H. Q. Xu, Giant, Level-Dependent $g$ Factors in InSb Nanowire Quantum Dots, \href{https://pubs.acs.org/doi/10.1021/nl901333a}{Nano. Lett. {\bf9}, 3151 (2009)}.
\bibitem{Mu2021}J. Mu, S. Huang, Z.-H. Liu, W. Li, J.-Y. Wang, D. Pan, G.-Y. Huang, Y. Chen, J. Zhao, and H. Q. Xu, A highly tunable quadruple quantum dot in a narrow bandgap semiconductor InAs nanowire, \href{https://pubs.rsc.org/en/content/articlehtml/2021/NR/D0NR08655J}{Nanoscale {\bf13}, 3983 (2021)}.
\bibitem{Jong2019}D. Jong, J. Veen, L. Binci, A. Singh, P. Krogstrup, L. P. Kouwenhoven, W. Pfaff, and J. D. Watson, Rapid Detection of Coherent Tunneling in an
InAs Nanowire Quantum Dot through Dispersive Gate Sensing, \href{https://journals.aps.org/prapplied/abstract/10.1103/PhysRevApplied.11.044061}{Phy. Rev. Applied {\bf11}, 044061 (2019)}.



\bibitem{com} Note that the effective magnetic field induced by the spin-orbit interaction is perpendicular to the chain. Consequently, the boundary conditions along the chain  involve the momentum (i.e., the derivative with respect to $x$)  and not the covariant momentum.







%

\end{thebibliography}
\end{document}